\pdfoutput=1
\documentclass[symmetry,article,accept,pdftex,moreauthors]{Definitions/mdpi}

\firstpage{1} 
\pubvolume{15}
\issuenum{11}
\articlenumber{1970}
\pubyear{2023}
\copyrightyear{2023}
\externaleditor{Preprint number of University of Zagreb: {\bf $\,$ ZTF-EP-23-04}
\linebreak
Preprint number of Ruđer Bo\v{s}kovi\'{c} Institute: {\bf $\,$ RBI-ThPhys-2023-27}}
\datereceived{3 July 2023 } 
\daterevised{11 September 2023 } % Comment out if no revised date
\dateaccepted{21 September 2023 } 
\datepublished{24 October 2023} 
\hreflink{https://doi.org/10.3390/\linebreak sym15111970} % If needed use \linebreak

\Title{Neutrino %MDPI: we removed ZTF-EP-23-04,  RBI-ThPhys-2023-27. please confirm.
 Oscillations
in Finite Time Path Out-of-Equilibrium Thermal Field Theory}
\TitleCitation{Neutrino Oscillations
in Finite Time Path Out-of-Equilibrium Thermal Field Theory}

\Author{{Ivan~}
Dadi\'c $^{1}$  and {Dubravko} Klabu\v car $^{2,}${*}\orcidA{}
}
\AuthorNames{I.~Dadi\'c and D. Klabu\v car}

\AuthorCitation{Dadi\'c, I.; Klabu\v car, D.}

\address{%
$^{1}$ \quad Ruđer Bo\v{s}kovi\'{c} Institute, Bijenička cesta 54, 10000 Zagreb,
                   Croatia; ivan.dadic@irb.hr\\
$^{2}$ \quad Physics Department, Faculty of Science, University of Zagreb,
Bijeni\v cka cesta 32, 10000 Zagreb, Croatia}

\corres{Correspondence: klabucar@phy.hr}

\usepackage{amssymb}
	\makeatother
\abstract{We demonstrate that the Finite-Time-Path Field Theory 
is an adequate tool for calculating neutrino oscillations.
We apply this theory using a mass-mixing Lagrangian which involves
the correct Dirac spin and chirality structure and a
Pontecorvo--Maki--Nakagawa--Sakata (PMNS)-like mixing matrix.
The model is exactly solvable. The Dyson--Schwinger equations
transform propagators of the input free (massless) flavor neutrinos 
into a linear combination of oscillating (massive) neutrinos. 
The results are consistent with the predictions of the PMNS matrix
while allowing for extrapolation to early times. }

       \def\kB {k\llap{/\kern1pt}}
        \def\pB {p\llap{/\kern1pt}}
        \def\qB {q\llap{/\kern1pt}}
        \def\KB {K\llap{/\kern1pt}}
        \def\PB {P\llap{/\kern1pt}}
        \def\QB {Q\llap{/\kern1pt}}
        \def\FB {F\llap{/\kern1pt}}
        \def\SigmaB {\Sigma\llap{/\kern1pt}}
        \def\nB {n\llap{/\kern1pt}}
        \def\deltaB{\delta\llap{/\kern1pt}}
	\def\partialB{\partial\llap{/\kern1pt}}

\begin{document}
\section{\bf~Introduction}

The evidence for neutrino oscillations,
which solve the solar neutrino problem~\cite{Davis:1968cp},
started with the study of atmospheric oscillations~\cite{fukuda:1998}. Evidence
has been collected from various sources and~for various distances from
the sources over~a wide range of neutrino energies and detectors~\cite{Appolonio:1999,Abazajian:2012ys,aguilar:2001,Schechter:1980gr,Eguchi:2003,
 Hosaka:2006,Cervera:2000,Barger:2002,Yasuda:2004,Burguet:2001,
Minakata:2001qm,Fogli:1996,Abe:2011,adamson:2011,abe:2012,an:2012,
ahn:2012,abe:2021,
NOvA:2021nfi,gonzales:2021,capozzi:2021,salas:2021,cleveland:1998,kaether:2010,
abdurashitov:2009,cravens:2008,abe:2011,nakajima:2020,aharmim:2013,
bellini:2011,bellini:2010,bellini:2004,gando:2013,abe:2018,aartsen:2015,
bezerra:2022,adey:2018,yoo:2020,adamson:2013,dunne:2020,himmel:2020}.

Neutrino oscillations, as~well as similar problems such as kaon
oscillations and decays, suffer from the absence of a proper formalism
for the features in which a finite-time description is~essential.

The most important aspect of this paper is to demonstrate that the 
Finite Time Path Field Theory (FTPFT) is an adequate
tool~\cite{Altherr:1994,Boyanovsky:2005} to treat such problems.
In particular, we start with the free Lagrangian ${\cal L}_{0}$
of massless neutrinos with three flavors; however, we mix them
dynamically through an interaction Lagrangian ${\cal L}_{I}$,
as in addition to the standard weak interaction term ${\cal L}_W$
it has the term ${\cal L}_{Mix}$, which contains a
Pontecorvo--Maki--Nakagawa--Sakata (PMNS)-like matrix involving
neutrino masses. The~term ${\cal L}_W$ is of course crucial 
for the processes of creation and detection of flavor neutrinos.
Nevertheless, we adopt the approximation where the neutrino mixing
is fully due to the term ${\cal L}_{Mix}$, that is, we neglect the
residual influence of ${\cal L}_W$ on the neutrino masses and mixing.
Of course, we must keep in mind the caveat that not calculating
these corrections misses the contributions from the vacuum
condensate, which Blasone~et~al. 
\cite{Blasone:2018hah,Blasone:2022joq,Blasone:2023pvq}
found to be important corrections to the PMNS result.
Nevertheless, they vanish in the relativistic limit,
while the PMNS result is recovered~\cite{Blasone:2018hah,Blasone:2022joq,Blasone:2023pvq}.
Because the case of nonrelativistic neutrino kinematics seems
to be far from present experimental capabilities, we proceed
to study neutrino propagators in our aforedescribed framework,
which provides us with an exactly solvable model suitable for
studying how FTPFT can be applied to neutrino~oscillations.

Through the exact solution of the Dyson--Schwinger equations (DSE),
we then obtain propagators (with the retarded $S_R$ and
advanced $S_A$ components) of oscillating (massive) neutrinos
which couple to the weak interactions.  The~DSE equation for
the Keldysh component ($S_K$) of neutrino propagators leads to
two possible histories distinguished by two of three masses
(say, $m_i$ and $m_j$) for which the interference leads to oscillations. 
An equal time limit of $S_K$ provides the number of neutrinos at time $t$.

The origin of neutrino masses  has not yet been answered conclusively.
 The variety of extant approaches includes the Majorana mass term, 
 right-handed neutrinos with a see-saw mechanism, supersymmetric 
 extensions of Standard model, and more. One intriguing possibility
 is the generation of neutrino masses and mixing via the aforementioned
 flavor vacuum condensate~\cite{Blasone:2018hah,Blasone:2022joq,Blasone:2023pvq}.
 However, in~the present paper we do not address the issue of the
 origin of neutrino masses, and focus on demonstrating the
 calculation of neutrino oscillations in the FTPFT~framework.

Whatever the mechanism of neutrino mass generation,
in this paper we assume that it manifests itself
through an additional term in the Lagrangian: the mass mixing term ${\cal L}_{Mix}$.
Here, the interaction Lagrangian contains the mass mixing matrix in a close analogy 
to the PMNS matrix except~with an explicit Dirac spin and chirality structure.
The model is exactly~solvable.

To calculate neutrino oscillations, we use FTPFT; 
 \cite{schwinger:1961,keldysh:1964} {in addition, see} 
\cite{kadanoff:1962,danielewicz:1984,
chou:1985,rammer:1986,landsman:1987,calzetta:1988,mlebellac:1996,
brown:1988,dadic:1999,blaizot:2002,Dadic:2001,Dadic:2002,Arleo:2004gn,
Dadic:2009zz,Millington:2012pf,
Millington:2013qpa,Dadic:2019mm,Dadic:2020}, often
called the closed time path (CTP) thermal field theory (TFT).
We find this to be an appropriate tool for calculating a wide variety
of problems, including equilibrium and nonequilibrium TFT problems, for which
it was originally constructed \cite{schwinger:1961,keldysh:1964},
as well as scattering processes, decays, and oscillations.

 How can field theory describe oscillations?
 It depends on which version of field theory is employed
 (for example, see~\cite{Beuthe:2001rc,Akhmedov:2010ms} and references therein).
If we look at the predominant field theory approach,
namely, the S-matrix approach, it exhibits excellent properties:
adiabatic switching of interaction, Gell-Mann--Low (G-L) theorem,
Fermi's golden rule, reduction technique, renormalization, {etc}.
However, for~time-dependent processes, notably oscillations,
it has nothing to say, in~the sense that it includes adiabatic
switching (on and off) of the interaction and~acts between the
infinite initial time $t_i = -\infty$ and final time $t_f =\infty$.

FTPFT, on~the other hand, uses a finite time path, no adiabatic switching,
and no G-L theorem. The~particle states are not exactly the eigenstates
of the full Hamiltonian. Renormalization, as~we have indicated
in~\cite{Dadic:2019mm,Dadic:2020}, is somewhat more involved;
nevertheless, it can be performed in finite time with
essentially the same technique of dimensional~regularization.

The drawback of this theory is the lack of a Gell-Mann--Low-type theorem.
Nevertheless, FTPFT does not contain disconnected subdiagrams owing to the
absence of the maximal-time vertices~\cite{Dadic:2001,Dadic:2002,Dadic:2009zz}.
Thus, this drawback may be cured after~carrying out all of the calculations in
a proper way by~considering the large time behavior of the contributions.
Then, in~this limit one can expect the G-L theorem to~again be valid.

The particle number is obtained as the equal time limit (ETL) of the re-summed $S_K$
propagator (or better) as~the average of ETL (AETL) for the contributions
distinguished by $t_{1f}-t_{2f}$ (see {Appendix} 
 \ref{appsec1}, especially {Appendix} 
\ref{FermionParticleNumber}).
These processes contribute inclusively to the particle number.
Fermi's golden rule need not be put in by hand, as it emerges
naturally during calculation of the convolution products ($*$-products).

In certain applications, the~solution of DSE can be written in a closed form.
The solution of DSE is illustrative:
conserving energy and momentum along the chain, 
the chain of retarded functions extends from the final time $t$
towards the earliest time $t_{min}\geq 0$. There, it meets
a $S_K$ propagator containing the initial distribution
function. At~that time, there is a (single) convolution product,
which does not conserve energy. Instead, it creates just enough 
oscillating deviations from the energy-conserving value to
satisfy the uncertainty relations. From~the lowest time, another
chain of advanced functions progresses back to the final time.
It conserves energy and momentum as well; however,~the energy of the
advanced chain differs from the energy of the retarded chain.
With the help of this energy difference, the~convolution
product creates oscillations. It is easy to see that
with increasing time the~uncertainty in energy becomes
smaller, which is in~the full agreement with the uncertainty~relations.

In certain special cases, energy is conserved at the remaining
$*$-product.
For instance, this is the case when Fermi's golden rule~emerges.

The remainder of this paper is organized as follows: first, we provide a short overview of
the PMNS-approach; then, we define the dynamics of our approach
through the mass mixing interaction Lagrangian ${\cal L}_{Mix}$
(\ref{LINT1}).
 
Our calculation is based on the DSE (\ref{Ads1n})--(\ref{Ads4n})
and their closed form solution. This solution defines the oscillating
neutrino which possesses flavor as massive and~weakly~interacting.
 
The equal time limit of the $S_K$ propagator is directly connected to
the neutrino number, as~it is measured at time $t$. In~the process
of calculation, we show how the number of complicated $*$-products
appearing in the formal solution simplifies to a single $*$-product,
which is necessary for the time evolution and energy--time uncertainty~relations.

As a result of this calculation, we obtain the neutrino number at time $t$.
Projection to the flavor degree of freedom exhibits oscillating behavior.
This result is consistent with Heisenberg's uncertainty relations between
time and energy. We reproduce the~PMNS result exactly within~the limit 
of ultrarelativistic neutrinos and~large time.

Finally, we discuss prospects for further~research.

Further details and calculation can be found in~{Appendix} 
 \ref{appsec1}.

\section{\bf PMNS Theory of Neutrino~Oscillation}

In the PMNS theory of neutrino oscillation~\cite{Pontecorvo:1957,Maki:1962,Wolfenstein:1978,Gribov:1969,Mikheyev:1985},
 three flavor neutrino states that interact weakly are mixed to three different superpositions of the neutrino states of definite mass (for mixing in gauge theories, see~\cite{Schechter:1980gr}). In~the flavor states, neutrinos are emitted and absorbed through weak processes; however,~they travel as mass~eigenstates.

This is mathematically {expressed as} 

\begin{equation}
 	\label{defrPMNS}
 \left|\nu_{\alpha}\right\rangle =
 \sum_{i }U_{\alpha i}\left|\nu_{i}\right\rangle \, ,  \qquad \qquad
 \left|\nu_{i}\right\rangle =
 	\sum_{\alpha}U^*_{\alpha i}\left|\nu_{\alpha} \right\rangle \, ,
 \end{equation}
where $\alpha = e,\mu,\tau$ are the  flavor indices, respectively, $e$-electron neutrinos,
$\mu$-muon neutrinos, and  $\tau $-tauon neutrinos, that label neutrinos with
definite flavors, while $m_{i},~~i=1,2,3,$ are the indices of the neutrino mass states.
To describe antineutrinos, it is necessary to use complex-conjugate matrices
$U_{\alpha i}\leftrightarrow U^*_{\alpha i}$. The matrix $U_{\alpha i}$ is
the PMNS-matrix, introduced in a way analogous to the CKM matrix describing
the mixing of~quarks.

The size of the matrix depends on theory. For~the standard three-neutrino theory,
the matrix is $3\times 3$. For~more neutrinos (including sterile ones),
the matrix could be larger. There are anomalies~\cite{Abazajian:2012ys,aguilar:2001,Aguilar:2021,mueller:2011,huber:2011,
giunti:2008,aartsen:2020,abe:2016,gonzales:2013,maltoni:2016,
guzzo:1991,roulet:1991,yasuda:2022}, suggesting that the model requires
 further~refinement.

The $3\times 3$ PMNS matrix (\ref{PMNS}), for~example, is provided
in~\cite{gonzales:2021}:\vspace{-5pt}
\begin{adjustwidth}{-\extralength}{0cm}%\centering
\begin{eqnarray}\label{PMNS} 
&&
U=\begin{pmatrix}U_{e_1}&U_{e_2}&U_{e_3}\cr
U_{\mu_1}&U_{\mu_2}&U_{\mu_3}\cr
U_{\tau_1}&U_{\tau_2}&U_{\tau_3}\cr\end{pmatrix}
% \cr\nonumber\\&&
% \cr
	\\&&
=\begin{pmatrix}1&0&0\cr
0&c_{23}&s_{23}\cr
0&-s_{23}&c_{23}\cr\end{pmatrix}
\begin{pmatrix}c_{13}&0&s_{13}e^{-i\delta }\cr
0&1&0\cr
-s_{13}e^{i\delta }&0&c_{13}\cr\end{pmatrix}
\begin{pmatrix}c_{12}&s_{12}&0\cr
-s_{12}&c_{12}&0\cr
0&0&1\cr\end{pmatrix}
\begin{pmatrix}e^{i\alpha _{1}/2}&0&0\cr
0&e^{i\alpha _{2}/2}&0\cr
0&0&1\cr\end{pmatrix},
\cr\nonumber
\end{eqnarray}
\end{adjustwidth}
where $ c_{ij} = \cos \theta_{ij}$ and~$s_{ij} = \sin \theta_{ij}$. The~phase factors 
$ \alpha_1 $ and $ \alpha_2 $ are relevant only if the neutrinos are Majorana particles,
i.e.,~if the neutrino is identical to its antineutrino, otherwise they 
can be ignored. The~phase factor $\delta$ measures the degree of violation of 
CP symmetry, which has not yet been observed~experimentally.

Note that Relation (\ref{defrPMNS}) is valid only approximately, as~particles
with the same energy and momentum but~different masses cannot simultaneously be 
on the mass shell. The~same feature appears in the usual treatment of neutrino oscillations 
in field theory, where it is known as mass-shell approximation~\cite{Beuthe:2001rc,Akhmedov:2010ms,Alves:2021rjc,Banks:2022gwq}.
 
A neutrino travels through space as a massive neutrino. Its time evolution is
described (in  units $c=1,\hbar =1$) by $\left|\nu _{i}(t)\right\rangle =
e^{-\,{\rm i}\,\left(E_{i}t-{\vec {p}}_{i}\cdot {\vec {x}}\right)}
\left|\nu _{i}(0)\right\rangle $ for each mass- and energy-eigenstate
$\left|\nu _{i}\right\rangle$ with mass $m_i$ and eigenenergy
$\, E_i = \sqrt{p_i^2 + m_i^2\, }$.
However, weak processes actually produce flavor states
$\left|\nu _{\alpha}\right\rangle$, which are, in~accordance with
Equation~(\ref{defrPMNS}), superpositions of these mass- and energy-eigenstates
$\left|\nu _{i}\right\rangle$. Due to their energy differences $E_i - E_j$,
such superpositions have an energy uncertainty $\Delta E$ (\ref{DeltaE}).

In the quantum mechanical treatment of neutrino oscillations, it is standard
to assume (for example,~see~\cite{Bilenky,Bilenky+al}) that all mass
eigenstates $\left|\nu_{i}\right\rangle$ have the same momentum, {i.e.},
their energies $E_i$ differ by their masses $m_{i}$ and~the flavor neutrino
state $\left|\nu_{\alpha}\right\rangle$ is determined by the~momentum.

In the ultrarelativistic limit, $ \left|{\vec {p}}_{i}\right|=p_{i}\gg m_{i}$.
Thus, $p_{i} \approx E$, the~neutrino energy in the limit $m_i \to 0$, such that
 $\, \forall \, i$:
\begin{equation}
E_{i}\simeq p_{i}+{\frac{m_{i}^{2}}{2p_{i}}}\approx E+{\frac{m_{i}^{2}}{2E}} \, , 
\qquad {\rm and} \qquad t \approx L \, ,
	\label{inULRelLim}
\end{equation}
where $t$ is the time from the beginning of evolution and
$L$ is the distance~traveled.

In the process of measurement, the neutrino is projected back to the flavor states.
The probability that the initial neutrino with flavor $\alpha $ will be detected later
 as having flavor $\beta $ is defined~as

$ P_{\alpha \rightarrow \beta } = 
\left|\left\langle \left.\nu _{\beta }(L)\right|\nu _{\alpha }\right\rangle \right|^{2} = 
\left|\sum _{i}U_{\alpha i}^{*}U_{\beta i}e^{-{\rm i}{\frac {m_{i}^{2}L}{2E}}}\right|^{2}.$

This can be written as
\begin{eqnarray}\label{PMNS1} 
&&
 P_{\alpha \rightarrow \beta }=\delta _{\alpha \beta }
-4\sum _{i>j}{\rm {Re}}\left(U_{\alpha i}^{*}U_{\beta i}U_{\alpha j}U_{\beta j}^{*}\right
)\sin ^{2}\left({\frac {\Delta m_{ij}^{2}L}{4E}}\right)
\cr\nonumber\\&&
+2\sum _{i>j}{\rm {Im}}\left(U_{\alpha i}^{*}U_{\beta i}U_{\alpha j}U_{\beta j}^{*}\right
)\sin \left({\frac {\Delta m_{ij}^{2}L}{2E}}\right),
\end{eqnarray}
where $ \Delta m_{ij}^{2} \equiv m_{i}^{2}-m_{j}^{2}$.

The second term is related to CP asymmetry:
$$ A_{CP}^{(\alpha \beta )}=P(\nu _{\alpha }\rightarrow \nu _{\beta })-P({\bar {\nu }}_{\alpha }\rightarrow {\bar {\nu }}_{\beta })=4\sum _{i>j}
{Im} \left(U_{\alpha_i}^{*}U_{\beta_i}U_{\alpha j}U_{\beta_j}^{*}\right)\sin \left({\frac {\Delta m_{ij}^{2}L}{2E}}\right) \, .$$
{With} 
 the help of the Jarlskog invariant
$ \quad
% \operatorname 
{Im} \left(U_{\alpha_i}U_{\beta_i}^{*}
U_{\alpha_j}^{*}U_{\beta_j}\right)
=J\sum_{\gamma ,k}\varepsilon_{\alpha \beta \gamma }\varepsilon_{ijk} \,$, 
{the CP} asymmetry  is expressed as
$$ A_{CP}^{(\alpha \beta )}=16J\sum _{\gamma }\varepsilon _{\alpha \beta \gamma }
\sin \left({\frac {\Delta m_{21}^{2}L}{4E}}\right)\sin \left({\frac {\Delta m_{32}^{2}L}{4E}}\right)\sin \left({\frac {\Delta m_{31}^{2}L}{4E}}\right) \, .$$
{Note} that CP asymmetry in neutrino oscillations has not yet been~observed.

Because
\begin{equation}
\label{DeltaE}
 \Delta E  \leq  \max_{i,j=1,2,3} (E_{i} - E_{j})\approx
 \frac {\, m_{i}^{2}-m_{j}^{2} \, }{2E} \, , 
\end{equation}
{Equation}~(\ref{PMNS1}) satisfies the time--energy uncertainty relation
$\Delta E \, \Delta t\gtrsim 1$ of Heisenberg (and of Mandelstam and Tamm,
as discussed in~\cite{Bilenky+al}) for~\begin{equation}
	t \gtrsim  \frac{ 2 E }{ \, \Delta m_{ij}^{2} \, } \, .
\label{t_min}
\end{equation}

{For} times shorter than (\ref{t_min}), the~uncertainty relations are
fulfilled with the help of neutrino production processes. Specifically, they
produce the flavor states $\, \left|\nu_{\alpha}\right\rangle$, which
are superpositions (\ref{defrPMNS}) of $\, \left|\nu_{i}\right\rangle$,
the eigenstates of neutrino masses $m_i$, which are so close that they
cannot be resolved in a time shorter than (\ref{t_min}) \cite{Bilenky+al}.

The ultrarelativistic limit applies to all currently observed neutrinos, 
as it is known that the differences of their squared masses are of
the order $10^{-4}$ eV$^2$ and their energies are at least 1 MeV.
Measured oscillation distances $L$ are on the order of~kilometers.

 \section{\bf   Neutrino Oscillation as a Dynamical~Process}
\unskip  

\subsection{The Mass Mixing Term in the Interaction~Lagrangian}  

In order to treat neutrino oscillation as a dynamical process, we start with  
the neutrino mixing interaction ${\cal L}_{Mix}$ of the Dirac spinor
Lagrangian, defined by:
\begin{eqnarray}\label{LINT}
&&
{\cal L}_0(x)=\sum_{\alpha}\bar \nu_{\alpha}(x)i\partialB
 \nu_{\alpha}(x)
% \cr %\nonumber
	\\&&
{\cal L}_I = {\cal L}_W+{\cal L}_{Mix}
 \end{eqnarray}
\begin{eqnarray}\label{LINT1} 
&&
%	\cr\nonumber\\&&
{\cal L}_{Mix}(x)=\sum_{\alpha,i}\bar \nu_{\alpha}(x)
U^*_{\alpha,i} M_{ij}U_{\beta,j}\nu_{\beta}(x) + antineutrinos
% \cr\nonumber
%	\\&&
 \end{eqnarray}
\begin{eqnarray}\label{LINT2} 
&&
 M=\begin{pmatrix}m_1&0&0\cr
0&m_2&0\cr
0&0&m_3\cr\end{pmatrix} \, ,
%\nonumber
 \end{eqnarray}
where $U_{\alpha,i}$ is a $3\times 3$ matrix analogous to the PMNS-matrix.
Here, $\nu_{\alpha}$ are neutrino spinor wave functions
for flavor $\alpha$ and $m_i$ represents the propagating neutrino masses.
The initial conditions are fixed through the neutrino distribution function
(including either the chirality or a handedness projection operator) which are
built in the $S_{f,K}$ component of the flavor neutrino propagator.
The massless  chiral ``flavor neutrinos'' (with propagators
$S_{f}(p)$ and weak interaction) enter the DSE. The
self-energy is identified from the next-to-lowest order to the DSE through 
the use of (\ref{LINT1}) and (\ref{LINT2}), and is simply the constant matrix
\begin{eqnarray}\label{SE}
&&
 \Sigma_{\alpha,\beta}=  \Sigma_{\alpha,\beta,R}=\Sigma_{\alpha,\beta,A}
= - \, U^*_{\alpha,i} M_{ij}U_{\beta,j},~~\Sigma_{\alpha,\beta,K}=0 
\end{eqnarray}

The DSE for fermions and their formal solutions are provided in
{Appendix} \ref{appsec1}, 
 specifically, (\ref{s})--(\ref{ds4}).
         
The DSEs for oscillating neutrinos are
\begin{eqnarray}\label{Ads1n}
&&
{\tilde S}_{\beta,\alpha,R}=S_{\beta,R}\delta_{\beta ,\alpha}
+iS_{\beta,R}*\Sigma_{\beta,\eta,R}*{\tilde S}_{\eta,\alpha,R}
%\cr\nonumber
	\\
\label{Ads2n}
	&&
{\tilde S}_{\beta,\alpha,A}=S_{\beta,A}\delta_{\beta ,\alpha}
+iS_{\beta,\eta,R}*\Sigma_{\eta,\zeta,A}*{\tilde S}_{\zeta,\alpha,A}
%\cr\nonumber
	\\
\label{Ads3n}
	&&
{\tilde S}_{\beta,\alpha,K}=S_{\beta,K}\delta_{\beta ,\alpha}
+i[S_{\beta,\eta,R}*\Sigma_{\eta,\zeta,R}*{\tilde S}_{\zeta,\alpha,K}
	%\cr\nonumber
	\\
\label{Ads4n}
	&&
+S_{\beta,\eta,K}*\Sigma_{\eta,\zeta,A}*{\tilde S}_{\zeta,\alpha,A}
+S_{\beta,\eta,R}*\Sigma_{\eta,\zeta,K}*{\tilde S}_{\zeta,\alpha,A}] \, ,
\end{eqnarray}
where $S$ denotes the lowest-order Green function and the
same symbol with a tilde (${\tilde S}$) denotes the
corresponding re-summed Green function, as~explained in 
{Appendix} 
 \ref{DSE_for_fermions}.

\subsection{ Solution of the Dyson--Schwinger Equations for Oscillating
 Neutrinos}
The formal solution for (\ref{Ads1n})--(\ref{Ads4n}) is
\begin{eqnarray}\label{Ads2o}
&&
{\tilde S}_{\beta,\alpha,R}=[1-iS_R*\Sigma_R]_{\beta,\eta}^{-1}*
 S_{\eta,\alpha,R}=S_{\beta,\eta,R}
 *[1-i\Sigma_R*S_R]_{\eta,\alpha}^{-1},
\cr\nonumber\\&&
{\tilde S}_{\beta,\alpha,A}=[1-iS_A*\Sigma_A]_{\beta,\eta}^{-1}
* S_{\eta,\alpha,A}=S_{\beta,\eta,A}
*[1-i\Sigma_A*S_A]_{\eta,\alpha}^{-1},
\cr\nonumber\\&&
{\tilde S}_{\beta,\alpha,K}
=-S_{\beta,\eta,K,A}
*(1-i\Sigma_A*S_A)_{\eta,\alpha}^{-1}
+(1-iS_R*\Sigma_R)_{\beta,\eta}^{-1}*S_{\eta,\alpha,K,R}
\cr\nonumber\\&&
+i(1-iS_R*\Sigma_R)_{\beta,\eta}^{-1}*[S_{\eta,\zeta,R}
*\Sigma_{\zeta,\omega,K}*S_{\omega,\theta,A}
-S_{\eta,\zeta,R}*\Sigma_{\zeta,\omega,R}*S_{\omega,\theta,K,A}
\cr\nonumber\\&&
+S_{\eta,\zeta,K,R}*\Sigma_{\zeta,\omega,A}*
S_{\omega,\theta,A}]*(1-i\Sigma_A*S_A)_{\theta,\alpha}^{-1}.
\end{eqnarray}
{The} solution simplifies owing to the following three reasons:
\begin{enumerate}
\item The self-energy is a simple matrix, not a retarded or advanced function;

\item The~$*$-products among the bare propagators turn to algebraic products
	except for the case where one factor is retarded (R or K{,}R) and the other is advanced (A or K{,}A);

\item The~matrix $U$ is~unitary.\end{enumerate}

Next, we calculate  $(1-i\Sigma_{R(A)}*S_{R(A)})_{\theta,\alpha}^{-1}$;
the matrix  inversion is simple, and we obtain
\begin{eqnarray}\label{sum-1}
&&
 (1-i\Sigma*S_{R})_{\beta,\alpha}^{-1}(p)=
 \sum_iU^*_{\beta,i}{p^2+m_i\pB\over p^2-m_i^2+ip_0\epsilon}U_{\alpha,i} \, .
  \end{eqnarray}

Then, we obtain the re-summed retarded propagator~component:

\begin{eqnarray}\label{sumR}
&&
\tilde{S}_{\beta,\alpha,R}(p)= (1-i\Sigma*S_{R})_{\beta,\eta}^{-1}
S_{\eta,\alpha,R}(p)=
 \sum_iU^*_{\beta,i}{-i(m_i+\pB)
 \over p^2-m_i^2+ip_0\epsilon}U_{\alpha,i}
  \end{eqnarray}
and the resummed advanced propagator~component

\begin{eqnarray}\label{sumA}
&&
\tilde{S}_{\beta,\alpha,A}(p)= (1-i\Sigma*S_{A})_{\beta,\eta}^{-1}
S_{\eta,\alpha,A}(p)= \sum_iU^*_{\beta,i}{-i(m_i+\pB)
\over p^2-m_i^2-ip_0\epsilon}U_{\alpha,i} \, .
  \end{eqnarray}

We now have an important conclusion. Relations (\ref{sumR}) and (\ref{sumA})
 express the re-summed flavor propagator through the linear combination
 of ``propagating'' neutrino propagators. This should be contrasted to
 the ``similar'' (mathematically ill-defined) relations between wave 
 functions of flavor and propagating~neutrinos.
  
Now, we turn to the $\tilde{S}_{\beta,\alpha,K}$ propagator. Insertion of 
(\ref{sum-1})--(\ref{sumA}) into (\ref{Ads2o}) provides us with
\begin{eqnarray}\label{sumK}
&&
  {\tilde S}_{\beta,\alpha,K}
=-\delta_{f,\alpha}\sum_iU^*_{\beta,i}U_{\alpha,i}
{i\over p^2-m_i^2-ip_0\epsilon}
\cr\nonumber\\&&
\times \{ [1-2n_{f}(\vec p)]
{p_0+|\vec p|\over 2|\vec p|}(\gamma_0|\vec p|-\vec \gamma \vec p)
%\cr\nonumber\\&&
-[1-2n_{\bar f}(-\vec p)]{p_0-|\vec p|\over 2|\vec p|}
(\gamma_0|\vec p|+\vec \gamma\vec p)\}{1-\gamma_5\over 2}
\cr\nonumber\\&&
+\delta_{\beta,f}\sum_iU^*_{\beta,i}U_{\alpha,i}
{i\over p^2-m_i^2+ip_0\epsilon}
 \cr\nonumber\\&&
 \times\{ [1-2n_{f}(\vec p)]
{p_0+|\vec p|\over 2|\vec p|}(\gamma_0|\vec p|-\vec \gamma \vec p)
%\cr\nonumber\\&&
-[1-2n_{\bar f}(-\vec p)]{p_0-|\vec p|\over 2|\vec p|}
(\gamma_0|\vec p|+\vec \gamma\vec p)\}{1-\gamma_5\over 2}
\cr\nonumber\\&&
+i\delta_{f,\alpha} \sum_{i,j}U^*_{\beta,i}{p^2+m_i\pB
\over (p^2-m_i^2+ip_0\epsilon)( p^2+ip_0\epsilon)}U_{\alpha,i}
% \cr\nonumber\\&&
 [-im_i(-i\pB)
  \cr\nonumber\\&&
 *\{ [1-2n_{f}(\vec p)]
{p_0+|\vec p|\over 2|\vec p|}(\gamma_0|\vec p|-\vec \gamma \vec p)
\cr\nonumber\\&&
-[1-2n_{\bar f}(-\vec p)]{p_0-|\vec p|\over 2|\vec p|}
(\gamma_0|\vec p|+\vec \gamma\vec p)\}{1-\gamma_5\over 2}
\cr\nonumber\\&&
 +\{ [1-2n_{f}(\vec p)]
{p_0+|\vec p|\over 2|\vec p|}(\gamma_0|\vec p|-\vec \gamma \vec p)
\cr\nonumber\\&&
-[1-2n_{\bar f}(-\vec p)]{p_0-|\vec p|\over 2|\vec p|}
(\gamma_0|\vec p|+\vec \gamma\vec p)\}{1-\gamma_5\over 2}]*im_j(-i\pB)]
\cr\nonumber\\&&
\times U^*_{\alpha,j}{p^2+m_j\pB\over (p^2-m_j^2-ip_0\epsilon)
(p^2-ip_0\epsilon)}U_{\beta,j} \, ,
\end{eqnarray}
where $f$ is the flavor of the initial neutrino~beam.

The resummed Keldysh component ${\tilde S}_{\beta,\alpha,K}$ of the neutrino
propagator consists of two contributions:
(a) terms without $*$-products (which we call algebraic) and
(b) the terms with a single $*$-product (which we call~convolutional).
       
This propagator carries the information which, after~the equal time limit,
provides the number and momentum distribution of all types of flavor 
neutrinos measured at time $t$.

\subsection{The $*$-Products and the Average of Equal Time~Limits }

 The remaining $*$-product in (\ref{sumK}) between the retarded and advanced functions is expressed (see Equation (\ref{ffff}) for details) as follows:
 \vspace{-8pt}
\begin{adjustwidth}{-\extralength}{0cm}%\centering
\begin{eqnarray}\label{Affff}
&&
C_{X_0}(p_0,\vec p)=\int dp_{01}dp_{02}
P_{X_0}(p_0,{p_{01}+p_{02}\over 2})
{i\over 2\pi}{e^{-iX_0(p_{01}-p_{02}+i\epsilon)}-1 
\over p_{01}-p_{02}+i\epsilon}
%\cr\nonumber\\&&
A_{\infty,R}(p_{01},\vec p)B_{\infty,A}(p_{02},\vec p),
\cr\nonumber\\&&
P_{X_0}(p_0,p'_0)={1\over \pi}\Theta(X_0){\sin\left( 2X_0(p_0-p'_0)\right)\over p_0-p'_0},
\end{eqnarray}
\end{adjustwidth}
where the label $\infty $ means that the values of $A$ and $B$ should be
taken at infinite time. After time $t$, the~number of oscillating neutrinos
is expressed through the average equal-time limit of the re-summed Keldysh
component ${\tilde S}_{\beta,\alpha,K}$ of the neutrino propagator
(see {Appendix} \ref{appsec1}):
\begin{eqnarray}\label{nupaff}
&&
 1-\langle \, N_{\beta,\vec p}(t)  \, \rangle  \,
\cr\nonumber\\&&
={1\over 2\pi}\,[\, \lim_{0<\Delta \rightarrow 0}
+\lim_{0>\Delta \rightarrow 0} \, ]\int dp_0  \,
	e^{-ip_0\Delta} \, Tr [ \, {\gamma_0\over 4}
	\tilde S_{\beta,\beta,K,t}(p){1-\gamma_5\over 2} \, ],
\end{eqnarray}
where $\Delta=s_{01}-s_{02}$ and $X_0=(s_{01}+s_{02})/2=t$.
In this expression, we have taken into account that the initial condition {(\ref{Lf})} 
contains only flavor neutrinos of type $f$,~not antineutrinos,
and we calculate number of flavor neutrinos of type $\beta $. The
equal time limit removes projector $P$ from the above $*$-product: \vspace{-8pt}
\begin{adjustwidth}{-\extralength}{0cm}%\centering
\begin{eqnarray}\label{Afff5}
%&&
{1\over \pi}\lim_{\Delta\rightarrow 0}\int \!
dp_0 \, e^{-i\Delta p_0} C_{s_{02=t}}(p_0,\vec p)
%\cr\nonumber\\&&
=\int dp_{01}dp_{02}
{i\over 2\pi}{e^{-it(p_{01}-p_{02})} -1
\over p_{01}-p_{02}}
%\cr\nonumber\\&&
A_{\infty,R}(p_{01},\vec p)B_{\infty,A}(p_{02},\vec p) . \, \, 
\end{eqnarray}
\end{adjustwidth}

\subsection{Contributions to Neutrino~Oscillation}

  To calculate neutrino oscillation, we start with chiral fermion number
    (\ref{nupaff}) and insert the solution of the Dyson--Schwinger
     Equation~(\ref{sumK}).

 Upon managing all  $*$-products appearing in (\ref{sumK}), we end 
 up with two types of\linebreak contributions:
 \begin{enumerate}
\item Contributions without any $*$-product. These contributions (the first pair in
(\ref{sumK})) are independent in time. They should reproduce the
initial (input) density of neutrinos. By~calculating geometrical series 
term by term, one would obtain the lowest order providing the input 
neutrino density. All the higher term would vanish, as~they are equal 
time limit of the product of two or more retarded functions (or 
 two or more advanced functions). The re-summed propagator is nonperturbative,
 and consequently the result renormalizes the initial density.
Nevertheless, we obtain the input neutrino~density.

\item Contributions containing $S_{\alpha,K}$ (the second pair in
(\ref{sumK})). These refer to the initial input of flavor neutrinos of type
 $\alpha $. They contribute to~oscillation. 
  \end{enumerate}

  \subsection{The Algebraic~Term}
    
        Note here that we have to decide whether we approach equal
         time from above or~from below. The~final result does not 
         depend on our choice; however,
        for our choice \linebreak $\Delta=s_{01}-s_{02}>0$ the~advanced 
        contribution should~vanish.
  
   We calculate the contribution of  $[1-iS_{R}\Sigma_{R}]^{-1}
    S_{\alpha,K,R}$ (at  $n_{\bar f}(-\vec p)=0$
   \vspace{-10pt}
\begin{adjustwidth}{-\extralength}{0cm}%\centering
\begin{eqnarray}\label{nupalg}
     &&
      I_{\alpha,alg}=1-\langle \,N_{\alpha, alg}(\vec p)(t)  \, \rangle  
      \, ={ i\over  \pi}
     \lim_{s_{01}\rightarrow s_{02}=t}\int dp_0 
     e^{-ip_0(s_{01}-s_{02})}
     \cr\nonumber\\&&
     {1\over 4}Tr \gamma_0
 \delta_{\alpha,f}\sum_iU^*_{\alpha,i}U_{\alpha,i}
{i\over p^2-m_i^2+ip_0\epsilon}
 %\cr\nonumber\\&&
  [1-2n_f(\vec p)]
{p_0+|\vec p|\over 2|\vec p|}(\gamma_0|\vec p|-\vec \gamma \vec p)
%\cr\nonumber\\&&
{1-\gamma_5\over 2}
    \cr\nonumber\\&& 
    {1\over 4}Tr\gamma_0(\theta+\eta\pB)
     (\gamma_0|\vec p|+\vec \gamma\vec p)
    [1-\gamma_5]=\theta|\vec p|
   \cr\nonumber\\&&
   I_{\alpha,alg}=
     \lim_{s_{01}\rightarrow s_{02}=t}\int dp_0 
     e^{-ip_0(s_{01}-s_{02})}
   %  \cr\nonumber\\&&
     \sum_i|U_{\alpha,i}|^2
{i\over p^2-m_i^2+ip_0\epsilon}
 %  \cr\nonumber\\&&
[1-2n_{f}(\vec p)]
    {p_0+|\vec p|\over 2\pi}
          \end{eqnarray}\end{adjustwidth}
       {Finally,} we find
\begin{eqnarray}\label{Anupalg1}
     &&
   I_{\alpha,alg}=\delta_{\alpha,f}[1-n_{f}(\vec p)].
      \end{eqnarray}

    This contribution  is the only time-independent contribution to neutrino 
    yield. As~we will see, it confirms the conservation of the total number of 
    neutrinos in our model. It indicates that the eventual 
    (finite) wave function
    renormalization is not~necessary.

 \subsection{$*$-Product, Term Containing $S_{K,R}$}
      
      In the terms with a convolution product, the~chirality 
      projector (${1-\gamma_5\over 2}$) appears twice: once to select chiral 
      neutrinos in initial distribution function, and~a second time to be
      selected by the weak interaction measurement~device:
     \begin{adjustwidth}{-\extralength}{0cm}%\centering
\begin{eqnarray}\label{K;R}
     &&
     I_{S_{conv,S_{K,R}}}
     =     {1\over 4}Tr \gamma_0
    { i\over 2\pi^2 }\int dp_{01}dp_{02}{e^{-i(p_{01}-p_{02})t}-1
    \over p_{01}-p_{02}} 
    \cr\nonumber\\&&
    i[1-i\Sigma S_{R}]^{-1}S_{K,R}\}(p_{01})
      \{\Sigma S_{A}[1-i\Sigma S_{A}]^{-1}\}(p_{02})  
     [1-\gamma_5]/2
       \cr\nonumber\\&&
       = \delta_{\alpha,f}\sum_{i,j}U^*_{\beta,i}U_{\alpha,i}U^*_{\alpha,j}U_{\beta,j}
       Tr \gamma_0
       { i\over 16\pi ^2 }\int dp_{01}dp_{02}{e^{-i(p_{01}-p_{02})t}-1
       \over p_{01}-p_{02}}
      % \cr\nonumber\\&&
i {p_1^2+m_i\pB_1
\over (p_1^2-m_i^2+ip_{01}\epsilon)( p_1^2+ip_{01}\epsilon)}
\cr\nonumber\\&&
 \{ [1-2n_f(\vec p)]
{p_{01}+|\vec p|\over 2|\vec p|}(\gamma_0|\vec p|-\vec \gamma \vec p)
%\cr\nonumber\\&&
-[1-2n_{\bar f}(-\vec p)]{p_{01}-|\vec p|\over 2|\vec p|}
(\gamma_0|\vec p|+\vec \gamma\vec p)\}[1-\gamma_5]m_j(-i\pB_2)]
\cr\nonumber\\&&
\times{p_2^2+m_j\pB_2\over (p_2^2-m_j^2-ip_{02}\epsilon)
(p_2^2-ip_{02}\epsilon)}{1-\gamma_5\over 2}
\cr\nonumber\\&&
  p_1=(p_{01},\vec p),~~ p_2=(p_{02},\vec p).
      \end{eqnarray}
\end{adjustwidth}

      The trace~is 
     \begin{eqnarray}\label{trK;R5}
   &&
   {1\over 4}Tr\gamma_0 (p^2_1+m_i \pB_1)
      [\gamma_0|\vec p|\mp\vec \gamma\vec p] [1- \gamma_5]
      (m_j+\pB_2)[1- \gamma_5]=2m_jp_1^2|\vec p|.
     \end{eqnarray}

    Now, we~have

\begin{eqnarray}\label{K;R2}
     &&
     I_{S_{conv,S_{K,R}}}
         = i\delta_{\alpha,f}\sum_{i,j}U^*_{\beta,i}U_{\alpha,i}U^*_{\alpha,j}U_{\beta,j}
       { im^2_j\over 4\pi ^2 }\int dp_{01}dp_{02}{e^{-i(p_{01}-p_{02})t}-1
       \over p_{01}-p_{02}}
       \cr\nonumber\\&&
{ \{ [1-2n_f(\vec p)](p_{01}+|\vec p|)
-[1-2n_{\bar f}(-\vec p)](p_{01}-|\vec p|)\}\over (p_1^2-m_i^2+ip_{01}\epsilon)(p_2^2-m_j^2-ip_{02}\epsilon)}.
%\cr\nonumber\\&&
      \end{eqnarray}

The integrals are performed by closing the integration path from below
       for $dp_{01}$ and~from above for $dp_{02}$.
        \vspace{-10pt}

\begin{adjustwidth}{-\extralength}{0cm}%\centering
\begin{eqnarray}\label{int1a}
     &&
    \int dp_{01}{e^{-i(p_{01}-p_{02})t}-1
    \over p_{01}-p_{02}}{p_{01}\pm|\vec p|\over
     p^2_1-m_{i}^2+i\epsilon p_{01}}
%      \cr\nonumber\\&&
    =-i\pi \sum_{\lambda_i=\pm 1}
    {(e^{-i(\lambda_i\omega_{i}-p_{02})t}-1)
    (\omega_{i}\pm\lambda_i|\vec p|)
    \over \omega_{i}(\lambda_i\omega_{i}-p_{02})}
       \cr\nonumber\\&&
    \int dp_{02}{e^{-i(\lambda_i\omega_{i}-p_{02})t}-1
    \over (\lambda_i\omega_{i}-p_{02})
    % \nonumber\\&&
    (p_2^2-m_{j}^2-ip_{02}\epsilon)}
   % \cr\nonumber\\&&
    =i\pi \sum_{\lambda_j=\pm 1}\lambda_j
    {e^{-i(\lambda_i\omega_{i}-\lambda_j\omega_{j})t}-1
    \over \omega_{j}( \lambda_i\omega_{i}-\lambda_j\omega_{j}) }
                 \end{eqnarray}
\end{adjustwidth}
     {Here,} $\omega_{i}=[|\vec p|^2+m^2_{i}]^{1/2}$ and
    $\omega_{j}=[|\vec p|^2+m^2_{j}]^{1/2}$

    Thus, the contribution consists of contributions from all singularities
    (four terms):
\begin{eqnarray}\label{ABK;R}
     &&
     I_{S_{conv,S_{K,R}}}
     = -{m_j^2\over 4} \delta_{\alpha,f} \sum_{ij}
      U^*_{\beta,i}U_{\alpha,i}U^*_{\alpha,j}U_{\beta,j}
       \sum_{\lambda_i,\lambda_j=\pm 1}\lambda_j
    {e^{-i(\lambda_i\omega_{i}-\lambda_j\omega_{j})t}-1
    \over \omega_{i}\omega_{j}
    ( \lambda_i\omega_{i}-\lambda_j\omega_{j}) }      
    \cr\nonumber\\&&
      [\lambda_i|\vec p|-n_{f}(\vec p)(\omega_{i}
      +\lambda_i|\vec p|)
%\cr\nonumber\\&&
+n_{\bar f}(-\vec p)] (\omega_{i}-\lambda_i|\vec p|)].
    \end{eqnarray}

    After setting $n_{-\alpha}(- \vec p)=0$, the~integral~is 
   \begin{eqnarray}\label{int1}
     &&
     I_{S_{conv,S_{K,R}}}
     = -{m_j^2\over 4} \delta_{\alpha,f} \sum_{ij}
      U^*_{\beta,i}U_{\alpha,i}U^*_{\alpha,j}U_{\beta,j}
       \sum_{\lambda_i,\lambda_j=\pm 1}\lambda_j
    {e^{-i(\lambda_i\omega_{i}-\lambda_j\omega_{j})t}-1
    \over \omega_{i}\omega_{j}
    ( \lambda_i\omega_{i}-\lambda_j\omega_{j}) }      
    \cr\nonumber\\&&
      [\lambda_i|\vec p|-n_{f}(\vec p)(\omega_{i}
      +\lambda_i|\vec p|)
       \cr\nonumber\\&&
       \omega_{i}=[|\vec p|^2+m^2_{i}]^{1/2},~~
    \omega_{j}=[|\vec p|^2+m^2_{j}]^{1/2}.
    \end{eqnarray}

  \subsection{$*$-Product,  Term Containing $ S_{K,A}$ and $i\leftrightarrow j$ Terms}
     
    This integral is similar to the previous~one: 
    \begin{eqnarray}\label{int2}
     &&
     I_{S_{conv,S_{K,A}}}
     = -{m_i^2\over 4} \delta_{\alpha,f} \sum_{ij}
      U^*_{\beta,i}U_{\alpha,i}U^*_{\alpha,j}U_{\beta,j}
        %\cr\nonumber\\&&
       \sum_{\lambda_i,\lambda_j=\pm 1}\lambda_i
    {e^{-i(\lambda_i\omega_{i}-\lambda_j\omega_{j})t}-1
    \over \omega_{i}\omega_{j}
    ( \lambda_i\omega_{i}-\lambda_j\omega_{j}) }    
     \cr\nonumber\\&&
      [\lambda_j|\vec p|-n_{f}(\vec p)
 (\omega_{j}+\lambda_j|\vec p|)].
  \end{eqnarray}

    With $i$ and $j$ interchanged, the contributions~are 
    \begin{eqnarray}\label{int1ji}
     &&
     I_{S_{conv,S_{K,R}}ji}
     = -{m_i^2\over 4} \delta_{\alpha,f} \sum_{ij}
      U^*_{\beta,j}U_{\alpha,j}U^*_{\alpha,i}U_{\beta,i}
        %\cr\nonumber\\&&
       \sum_{\lambda_i,\lambda_j=\pm 1}\lambda_i
    {e^{-i(\lambda_j\omega_{j}-\lambda_i\omega_{i})t}-1
    \over \omega_{i}\omega_{j}
    ( \lambda_j\omega_{j}-\lambda_i\omega_{i}) }    
     \cr\nonumber\\&&
     [\lambda_j|\vec p|-n_{f}(\vec p)
 (\omega_{ki}+\lambda_i|\vec p|)]
  \end{eqnarray}
    and 
 \begin{eqnarray}\label{int2ji}
     &&
     I_{S_{conv,S_{K,A}ji}}
     = -{m_j^2\over 4} \delta_{\alpha,f} \sum_{ij}
      U^*_{\beta,j}U_{\alpha,j}U^*_{\alpha,i}U_{\beta,i}
      %  \cr\nonumber\\&&
       \sum_{\lambda_i,\lambda_j=\pm 1}\lambda_j
    {e^{-i(\lambda_j\omega_{j}-\lambda_i\omega_{i})t}-1
    \over \omega_{i}\omega_{j}
    ( \lambda_j\omega_{j}-\lambda_i\omega_{i}) }    
    \cr\nonumber\\&&
    [ \lambda_i|\vec p|-n_{f}(\vec p)
 (\omega_{ki}+\lambda_i|\vec p|)].
  \end{eqnarray}
    %  
         
     %\newpage
\subsection{The Dominant~Contribution}
\label{TheDominantContribution}

The results of the previous section are consistent with Heisenberg's 
uncertainty condition between energy and time.
In our further calculations, we assume that the time is large compared to the energy differences
$|\omega_i-\omega_j|$. Together with the fact that the 
measured neutrinos $(m_{i}<<|\vec p|)$ are mostly ultrarelativistic,
with $|\vec p|$ ranging from 1 MeV to 10 MeV, the~following
approximations are justified:\vspace{-3pt}
\begin{adjustwidth}{-\extralength}{0cm}%\centering
\begin{equation}
\omega_{i}-\omega_{j}\approx {m^2_{i}-m^2_{j}\over 2|\vec p|}, 
\qquad
\omega_{i}+\omega_{j}\approx 2|\vec p|, \qquad 
\omega_{i}+|\vec p|\approx 2|\vec p| \qquad {\rm and} 
\qquad \omega_{i}-|\vec p|\approx {m^2_{i}\over 2|\vec p|} \, . \,
	\label{ultrarel_approxs}
\end{equation}
\end{adjustwidth}
{Owing} to the above, all contributions proportional to 
$[1-2n_{f}(\vec p)]$ are dominated by
\linebreak $\lambda_i=\lambda_j=+1$, while the contributions proportional 
to $[1-2n_{\alpha,-}(-\vec p)]$ are dominated by 
$\lambda_i=\lambda_j=-1$. The~contribution from the constant
$1$ in $[1-2n_{f}]$ is killed by \mbox{this~procedure}.

Notice, however, that for short times ($t\leq{h\over 4\pi|\omega_i-\omega_j|},~i\neq j$)
the approximations are not justified, and it is necessary to deal with the full expression.
in order to satisfy Heisenberg's uncertainty relations among time and energy.

The above contributions then become
\begin{eqnarray}\label{Aint1}
     &&
     I_{S_{conv,S_{K,R}}d}        
     = m_j^2 \delta_{\alpha,f} \sum_{ij}
      U^*_{\beta,i}U_{\alpha,i}U^*_{\alpha,j}U_{\beta,j}
      %\cr\nonumber\\&&
    {e^{-i{m^2_{i}-m^2_{j}\over 2|\vec p|}t}-1
    \over  m^2_{i}-m^2_{j}}    
     %\cr\nonumber\\&&
     n_{f}(\vec p)
\end{eqnarray}
\begin{eqnarray}\label{Aint2}
     &&
     I_{S_{conv,S_{K,A}}d}
     = m_i^2 \delta_{\alpha,f} \sum_{ij}
      U^*_{\beta,i}U_{\alpha,i}U^*_{\alpha,j}U_{\beta,j}
      %\cr\nonumber\\&&
      {e^{-i{m^2_{i}-m^2_{j}\over 2|\vec p|}t}-1
    \over  m^2_{i}-m^2_{j}}
     n_{f}(\vec p)
\end{eqnarray}
\begin{eqnarray}\label{int1jid}
     &&
     I_{S_{conv,S_{K,R}}jid}
     =m _i^2 \delta_{\alpha,f} \sum_{ij}
      U^*_{\beta,j}U_{\alpha,j}U^*_{\alpha,i}U_{\beta,i}
         %  \cr\nonumber\\&&
                {e^{i{m^2_{i}-m^2_{j}\over 2|\vec p|}t}-1
    \over m^2_{i}-m^2_{j}}    
     n_{f}(\vec p)
  \end{eqnarray}
    and 
 \begin{eqnarray}\label{int2jid}
     &&
     I_{S_{conv,S_{K,A}jid}}
     = m_j^2 \delta_{\alpha,f} \sum_{ij}
      U^*_{\beta,j}U_{\alpha,j}U^*_{\alpha,i}U_{\beta,i}
       %\cr\nonumber\\&&
       {e^{i{m^2_{i}-m^2_{j}\over 2|\vec p|}t}-1
    \over m^2_{i}-m^2_{j}} 
  \, \,  n_{f}(\vec p) \, .
    \end{eqnarray}

By adding the contributions, we obtain
\begin{eqnarray}\label{int3}
  %   &&
I_{S_{conv,S_{K}}} = \,
 I_{S_{conv,S_{K,R}}d} \, - \, I_{S_{conv,S_{K,A}}d} \, + \,   
   I_{S_{conv,S_{K,R}}jid} \, - \, I_{S_{conv,S_{K,A}}jid}
\qquad	\qquad    \qquad &&
      \cr\nonumber\\ % &&
	      =\delta_{\alpha,f}  \sum_{i\leq j}
\{U^*_{\beta,i}U_{\alpha,i}U^*_{\alpha,j}U_{\beta,j}
       [e^{-i{m^2_{i}-m^2_{j}\over 2|\vec p|}t}-1]
       - [ U^*_{\beta,j}U_{\alpha,j}U^*_{\alpha,i}U_{\beta,i}
       e^{i{m^2_{i}-m^2_{j}\over 2|\vec p|}t}-1]\}
  % %\cr\nonumber\\&&
\,   n_{f}(\vec p)  \, , &&
     \end{eqnarray}
which can be further written as
\vspace{-6pt}\begin{adjustwidth}{-\extralength}{0cm}%\centering
\begin{eqnarray}\label{int4}
     &&
I_{S_{conv,S_{K}}} \, = \, - \, n_{f}\, (\vec p)
\, \sum_{i\leq j}  \qquad
  \cr\nonumber\\
&&  
\delta_{\alpha,f} n_{f}(\vec p)  
[\, - \,4Re(U^*_{\beta,i}U_{\alpha,i}U^*_{\alpha,j}U_{\beta,j})
       \sin^2{m^2_{i}-m^2_{j}\over 4|\vec p|} \, t
    + 2 \, Im(U^*_{\beta,j}U_{\alpha,j}U^*_{\alpha,i}U_{\beta,i})
       \sin{m^2_{i}-m^2_{j}\over 2|\vec p|}t \, ] \, . \qquad
       \end{eqnarray}
\end{adjustwidth}

   \section{\bf Final~Result}

From the preceding section, notably Equation~(\ref{nupaff}), and~from {Appendix} \ref{appsec1}, 
notably {Appendix} 
 \ref{A8}, it can be seen that
at time $t$ the~total number of particles of the flavor $\beta$
stemming from the initial flavor $\alpha$ is \vspace{-12pt}
\begin{adjustwidth}{-\extralength}{0cm}%\centering
\begin{eqnarray}\label{int7}
\langle \,N_{\beta,f, \vec p}(t)  \, \rangle  = 
	    \delta_{\alpha,f}n_{\alpha,+}(\vec p)
     % \nonumber\\&&   
	    & + & n_{f}(\vec p)
 \delta_{\alpha,f}\sum_{i\leq j} [ \, - \,4Re(U^*_{\beta,i}U_{\alpha,i}U^*_{\alpha,j}U_{\beta,j})
       \sin^2{m^2_{ki}-m^2_{lj}\over 4|\vec p|} \, t \,
  \cr\nonumber\\  
	    &+& 2Im(U^*_{\beta,j}U_{\alpha,j}U^*_{\alpha,i}U_{\beta,i})
       \sin{m^2_{i}-m^2_{j}\over 2|\vec p|}\, t \, ] \, .
      \end{eqnarray}

\end{adjustwidth}
Let us point out a few features of this expression:
\begin{enumerate}
\item The~result (\ref{int7}) is identical to the standard PMNS expression (\ref{PMNS1}).
The ultrarelativistic relation (\ref{inULRelLim}) reveals the equality of the arguments
of the sines, while division by the initial distribution of the number of particles
$n_{f}$ recasts (\ref{int7}) in terms of probability, as in (\ref{PMNS1}).
Thus, with~the same presently available inputs, our result (\ref{int7})
would provide the same numerical results as, for~example,~\cite{yasuda:2022}.

\item  If~we sum over $\beta$, the~oscillating contribution vanishes. This reflects 
the fact that the total neutrino number is conserved within the realm 
of chiral neutrinos; notably, the sterile neutrinos are not involved!
   Notice that our conclusion is valid for low energy neutrino beams as well;
   this is easily verified by looking at (\ref{int1})--(\ref{int2ji}). 
 
\item  The~results are valid for moderate energies, although~it is necessary to
  skip the simplifications in (\ref{inULRelLim}) and (\ref{ultrarel_approxs})
and include all~contributions.
\end{enumerate}

  \subsection*{ {Conclusions} 
}
  
In this paper, we apply the Finite Time Path Field Theory (FTPFT), originally
designed to deal with out-of-equilibrium many-body statistical ensembles,
to the problem in particle physics. We demonstrate that FTPFT
is an appropriate tool for the treatment of neutrino~oscillations.
  
We calculate neutrino oscillation within a simple model, with the interaction Lagrangian containing the term
        ${\cal L}_{Mix}$
built as mass mixing through the PMNS-matrix with~a built-in Dirac spinor and chirality
structure. The~model is exactly solvable.
This is an extension of standard PMNS-case, as it involves sterile neutrinos,
at least those with the same mass as flavor and propagating neutrinos. Of~course,
more sophisticated sterile neutrinos would require additional model~building.

The result is consistent with Heisenberg's uncertainty relation
between energy and time.
The flavor neutrinos are chiral spin $1/2$ particles, though a~massive neutrino
propagator should additionally contain the right-handed component along with the
re-summed oscillating propagator. Nevertheless, the~result is chirally
invariant, as it is dictated from the weak interaction Lagrangian
(${\cal L}_W$) taken into account through the factors $(1-\gamma_5)$
appearing at the beginning (production) and end (detection) stages.
The sterile neutrinos do not contribute to oscillations, even for
low energy beams. Our result coincides with the~PMNS-formula 
for~large times and for the ultrarelativistic case.

In the derivation of our result, the~PMNS relation {(\ref{defrPMNS})} 
 is not used.
 Instead of using the relation which mixes flavor and propagating
 neutrino states (and~in which at least one state cannot be
 on mass shell), the~mass mixing Lagrangian through DSE provides an
 equally significant relation among the oscillating neutrino
 propagators obtained from DSE {(\ref{sumR})} and {(\ref{sumA})}. They enable
 the calculation of the Keldysh component of the propagator
 $\tilde{S}_{\beta,\alpha,K}$ {(\ref{sumK})}, which by application
 of AETL provides measurable particle numbers {(\ref{int7})}.

The FTPFT approach successfully passes the test of neutrino oscillations,
in the sense that we demonstrate how to perform calculations of these
oscillations using the presented approach. We reproduce the standard PMNS results
in the present model, which introduces the neutrino masses precisely
through the PMNS-matrix in the mixing part of the model Lagrangian.
This means that while the~present approach does not address the issue of the
origin of neutrino masses, and~in general does not provide answers
about the dynamics and physics of processes, the existing
knowledge can be used as an input in the form of masses, matrices, self-energies,
{etc}. 

The application of the FTPFT approach to neutrino oscillations shows
that it is an interesting candidate for a complementary tool to the
S-matrix, which is formulated for infinite times and involves switching
interactions on and off adiabatically. In~the case of phenomena where
finite times are essential, such as the presently pertinent neutrino
oscillations as well as the oscillation of kaons and B and D mesons,
decays, and symmetry violations, it has been necessary to use largely heuristic
methods, such as elementary quantum mechanics in the PMNS approach and
the Gell-Mann--Pais approach for kaons. In~such cases, the~FTPFT
approach is obviously a candidate for a more rigorous~description.

Further tests of the FTPFT approach could proceed along the following main lines:
  \begin{enumerate}
\item Improving the model by taking into
    account the eventually-confirmed~anomalies.

\item If suitable for~considering decays of heavier neutrinos,
    the model could be easily adapted to build these features in.
    In this case, the~self-energies should again be provided through an
    adequate~calculation.
  
\item Applying the formalism to other oscillating and decay processes
    (e,g., decays of $K^0$, $D^0$, and $B^0$ mesons, 
positronium, the Cabbibo angle,etc.).
This work is in progress and almost completed. It roughly confirms
the Gell-Mann--Pais results. The~factor which limits the predictive power
is the rudimentary knowledge of the self-energies in the existing literature. 
Authors have mostly been
concerned with obtaining the imaginary parts (decay rates), while the real parts (mass shifts)
 often involve~renormalization.

\item In~a classical out-of-equilibrium problem, the~damping rates are the
    first thing to address. Braaten--Pisarski re-summation has provided a good
    start. Even for this case, two-loop self-energy diagrams contain minimal
    time vertices, and~possible ``upgrades'' could be very tricky.
    This is another area where work is in~progress.
  \end{enumerate}

\appendixtitles{no} 
\appendixstart
\appendix
\section[\appendixname~\thesection]{}\label{appsec1}
\appendixtitles{yes} 
\subsection[\appendixname~\thesubsection]{{Finite Time Path} 
 Field~Theory}

 To calculate neutrino oscillations, we use Finite Time Path Field Theory
 (FTPFT), often called the closed time path thermal field theory (CTP-TFT). 
 ``Thermal'' is used here for historical reasons, as the theory was first constructed to treat
 ensembles at the thermal equilibrium or~very close to it. Now, we find
 it to be an appropriate tool for calculation of a wide variety of problems, including
 equilibrium and nonequilibrium TFT problems (within the linear response
 approximation) as well as~scattering processes, decays, and oscillations.
There are a number of specific features distinguishing it from S-matrix theory:
  \begin{enumerate}
\item {The time} 
 path $C$ is closed and finite: $C=(0+i\epsilon,t+i\epsilon)U(t+i\epsilon,t-i\epsilon)U(t-i\epsilon,0-i\epsilon)$.

\item The subject of the S-matrix is amplitude (``wave function''), 
while in FTPFT it is a two-point~function.

\item The product of two point functions is not algebraic, instead being a 
convolution product (see {Appendix} 
 \ref{appsec2}); only under
special conditions does it become an algebraic~product.

\item Instead of Feynman propagators, matrix propagators
$S_{ij}, ~i=1,2,~j=1,2$ are obtained. These are linearly transformed into
$S_R$ (retarded), $S_A$ (advanced), and $S_K$ (Keldysh) propagators.
Our method further separates $S_K$ into its retarded and advanced pieces
($S_K=S_{K,R}-S_{K,A}$);
$S_K$ contains single particle distribution functions of the 
unperturbed system (i.e.,~as they are determined at $t=0$).

\item A~measured quantity is obtained as an equal time limit of $S_K$.
Compared to a scattering matrix, these measured quantities are more
inclusive: one particle is separated (and measured), while the
others are integrated over.
This is equivalent to the exclusive S-matrix approach. In~addition, the
calculated quantities correspond to a yield, i.e.,~the number
of particles found at time $t$, while the equivalent in the S-matrix approach is the
cross-section, i.e.,~related to the time derivative of the~yield. 

\item While primarily developed for thermal equilibrium and out-of-equilibrium
(particularly ``almost equilibrated'') ensembles, nothing prevents it 
from being applied to decays and oscillations (as in this paper), or
to scattering~processes.

\item For~application to scattering processes, it is necessary to choose initial 
single particle distributions (i.e., for incoming particles) as two plane 
waves of extremely low intensities. After calculation it is then necessary 
to carry out the $t\rightarrow \infty$ limit. In~our experience, the results are
physically equivalent to S-matrix calculation. Adiabatic switching 
(on and off) of the interaction is not possible in finite time. With an infinite 
time limit ($t\rightarrow \infty $), the lack of adiabatic switching 
does not~matter.
  \end{enumerate}
\subsection{Convolution Product of Two Two-Point~Functions }\label{appsec2}

The convolution product~\cite{Dadic:2001} of two Green functions 
is defined as
\begin{eqnarray}\label{pgf}
C=A*B \Leftrightarrow C(x,y)=\int dz A(x,z)B(z,y).
\end{eqnarray}
{In} terms of Wigner transforms (see \cite{Dadic:2001} for
more details), it becomes
\begin{adjustwidth}{-\extralength}{0cm}%\centering
\begin{equation}\label{ffff}
C_{X_0}(p_0,\vec p) \, = \,\int dp_{01} \, dp_{02} \,
P_{X_0}(p_0,{p_{01}+p_{02}\over 2}) \,
{i\over 2\pi} \, {e^{-iX_0(p_{01}-p_{02}+i\epsilon)} 
\over p_{01}-p_{02}+i\epsilon} \, 
A_{\infty}(p_{01},\vec p)B_{\infty}(p_{02},\vec p) \, ,
\end{equation}
\end{adjustwidth}
where the projector $P_{X_0}$ is
\begin{equation}\label{projector}
P_{X_0}(p_0,p'_0)={1\over 2\pi}
\Theta(X_0)\int_{-2X_0}^{2X_0}ds_0e^{is_0(p_0-p'_0)}
={1\over \pi}\Theta(X_0){\sin\left( 2X_0(p_0-p'_0)\right)\over p_0-p'_0}
\end{equation}
and
\begin{eqnarray}\label{isftf}
e^{-is_0p'_0}\Theta(X_0)\Theta(2X_{0}+s_{0})\Theta(2X_{0}-s_{0})
=\int dp_0e^{-is_0p_0}P_{X_0}(p_0,p'_0).
\end{eqnarray}
{It} is important to note that
\begin{eqnarray}\label{dsftf}
\lim_{X_0\rightarrow \infty}P_{X_0}(p_0,p'_0)
=\lim_{X_0\rightarrow \infty}{1\over \pi}
{\sin\left( 2X_0(p_0-p'_0)\right)\over p_0-p'_0}
=\delta(p_0-p'_0).
\end{eqnarray}
{The} retarded (advanced) function is supposed to satisfy
the following properties~\cite{Dadic:2001}:
\begin{enumerate}
\item[(1)] The function of $p_0$ is analytic above (below) the real axis;

\item[(2)] The function vanishes as $|p_0|$ approaches infinity in
the upper (lower) semiplane.
\end{enumerate}

If $A$ is an advanced projected operator, we can integrate the expression (\ref{ffff})
even further. After~closing the $p_{01}$ integration contour in the lower
semi-plane, we obtain
\begin{eqnarray}\label{fff4}
C_{X_0}(p_0,\vec p)=\int dp_{01}P_{X_0}(p_0,p_{01})
A_{\infty}(p_{01},\vec p)B_{\infty}(p_{01},\vec p).
\end{eqnarray}

{If} $B$ is a retarded projected operator, we can achieve the same result
by closing the $p_{02}$ integration contour in the upper~semi-plane.

In the $X_0 \to \infty $ limit (\ref{dsftf}), Equation~(\ref{fff4}) becomes
the simple algebraic product
\begin{eqnarray}\label{fff4i}
C_{\infty}(p_0,\vec p)=
A_{\infty}(p_{0},\vec p)B_{\infty}(p_{0},\vec p).
\end{eqnarray}

{Note} that Equations~(\ref{fff4}) and (\ref{fff4i}) are valid for combinations
$A_AB_A,~A_AB_R,~A_RB_R$, but~not for $A_RB_A$.

For $A_R$ (retarded) and $B_A$ (advanced), the~Equal Time Limit
(ETL) of the convolution product is obtained as follows:
\begin{adjustwidth}{-\extralength}{0cm}%\centering
\begin{equation}\label{fff5}
{1\over \pi}\lim_{\Delta\rightarrow 0}\int dp_0e^{-i\Delta p_0}C_{X_0}(p_0,\vec p)
	=\int dp_{01}dp_{02}
{i\over 2\pi}{e^{-iX_0(p_{01}-p_{02})} -1
\over p_{01}-p_{02}}
A_{\infty,R}(p_{01},\vec p)B_{\infty,A}(p_{02},\vec p)
\end{equation}
\end{adjustwidth}
where $\Delta=s_{01}-s_{02}$ and $\, X_0=(s_{01}+s_{02})/2$.
The ETL is finally changed to AETL (the average of ETL), where
$\lim_{\Delta\rightarrow 0}$ is replaced by 
${1\over2 }[\lim_{0<\Delta\rightarrow 0}+\lim_{0>\Delta\rightarrow 0}]$.
From the numerator, the~constant term ($-1$) could have been subtracted,
because for this term it is possible to close the integration path $dp_{01}$ from below
(or to close the integration path $dp_{02}$ from above), in order to~find
that the contribution vanishes. The~resulting ``kernel'' is not singular, and we omit the $i\epsilon $ prescription.

From now on, we skip the index $_{\infty}$ as~self-understandable.

\subsection{Massive Neutrino~Propagator}

The massive  neutrino propagators
with masses $m_i\neq 0, i=1,2,3$ for 
the case when the fermion and antifermion distribution are equal:

$n_+(\omega_p,\vec p)=n_-(\omega_p,-\vec p)$ is simple,
\begin{eqnarray}\label{RAKS}
&&D_{R}(p) \, = \, (\pB+m)\, G_{R}(p,m)~,
\cr\nonumber\\
&&D_{A}(p) \, = \, (\pB+m)\, G_{A}(p,m)~,
\cr\nonumber\\
&&
G_{R(A)}(p,m) \, = \, {-i\over p^2-m^2\pm 2ip_0\epsilon}~,
\cr\nonumber\\
&&D_{K}(p)
%\cr\nonumber\\&&
=D_{K,R}(p)-D_{K,A}(p),
\cr\nonumber\\&&
D_{K,R}(p)=-[1-2n(\omega_p)] \, (\PB+ {mp_0\over\omega_p}) \, G_{R}(p,m),
\cr\nonumber\\&&
D_{K,A}(p)=-[1-2n(\omega_p)] \, (\PB+ {mp_0\over\omega_p}) \, G_{A}(p,m),
\cr\nonumber\\&&
\pB=\gamma^{\mu}p_{\mu},~p=(p_0,\vec p),
~\PB=\gamma^{\mu}P_{\mu},~P=(\omega_p,{p_0\over \omega_p}\vec p),
~~%\cr\nonumber\\&&
\omega_p=\sqrt{\vec p^2+m^2 \, },
\end{eqnarray}
where $n(\omega_p)$ is the initial fermion distribution~function.

\subsection{``Flavor Neutrino''~Propagator}
    
 Flavor neutrinos $\nu_{e}$,  $\nu_{\mu}$, and $\nu_{\tau}$ are,
 by assumption, massless and chiral; only left-handed flavor neutrinos
 and right-handed antineutrinos
 exist:
\begin{eqnarray}\label{RAKSf}
&&S_{f,R}(p) \, = \, \pB\, G_{R}(p,0)~,
\cr\nonumber\\
&&S_{f,A}(p) \, = \, \pB\, G_{A}(p,0)~,
\cr\nonumber\\
&&
G_{R(A)}(p,0) \, = \, {-i\over p^2\pm 2ip_0\epsilon}~,
\cr\nonumber\\&&
\pB=\gamma^{\mu}p_{\mu},~p=(p_0,\vec p),
	~\PB=\gamma^{\mu}P_{\mu},~P=(|\vec p|,{p_0\over |\vec p|}\vec p) \, .
\end{eqnarray}

For unequal flavor neutrino and flavor antineutrino distributions, we have
\begin{eqnarray}\label{fne}
&&
n(p_0,\vec p)={1-\gamma_5\over 2}\Theta(p_0)
n_+(\vec p)-{1-\gamma_5\over 2}\Theta(-p_0)
n_-(-\vec p) \, ,
%\cr\nonumber\\&&
% ~~\nB=\pB/|\vec p|
\end{eqnarray}
where  $n (p_0,\vec p)$ is now a 4 $\times$ 4~matrix.

Now, we decompose Keldysh propagator into its retarded and advanced parts:
\begin{eqnarray}\label{SRAf}
&&S_{f,K}(p) \, = \, 
S_{f,K,R}(p)-S_{f,K,A}(p)~,
\cr\nonumber\\&&
S_{f,K,R}(p) \, = \, - \, G_{R}(p,0) \, L_f(p_0, \vec p)~,
\cr\nonumber\\&&
S_{f,K,A}(p) \, = \, - \, G_{A}(p,0) \, L_f(p_0, \vec p)~,
\end{eqnarray}
where
\begin{eqnarray}\label{Lf}
	L_f(p_0, \vec p) & = & [1-2n_{f}(\vec p)]
{p_0+|\vec p|\over 2|\vec p|}(\gamma_0|\vec p|-\vec \gamma \vec p){1-\gamma_5\over 2}
\cr\nonumber\\
	& - &[1-2n_{\bar f}(-\vec p)]{p_0-|\vec p|\over 2|\vec p|}
(\gamma_0|\vec p|+\vec \gamma\vec p){1-\gamma_5\over 2}.
\end{eqnarray}

We define the propagator for flavor antineutrinos 
by $\bar {\cal S}$, which is obtained from  $S$
via the replacement of $n_f \leftrightarrow n_{\bar f}$. 
These propagators satisfy the following properties under inversion:
\begin{equation*}
S_{f,R}(-p)=-\bar {\cal S}_{f,A}(p),
~~
S_{f,K,R}(-p)=-\bar {\cal S}_{f,K,A}(p).
\end{equation*}

    \subsection{Oscillating Neutrino~Propagator}

In~addition to flavor and propagating neutrinos, we have
defined ``oscillating neutrinos'' above by solving
Equations~(\ref{Ads1n})--(\ref{Ads4n})
through partial re-summation of DSE
over the powers of the mixing mass interaction self-energies.

  % \newpage

 \subsection{Dyson--Schwinger Equation for~Fermions}
\label{DSE_for_fermions}
  Here, $S$ is the lowest order Green function and
${\tilde S}$ is the re-summed Green function:
\begin{eqnarray}\label{ds}
[S^{-1}-i\Sigma]*{\tilde S}=1 ,
\end{eqnarray}
where
\begin{eqnarray}\label{s}
&&
S=\begin{pmatrix}S_R &S_K\cr 0 &S_A\cr\end{pmatrix},~~  
\Sigma=\begin{pmatrix}\Sigma_R &\Sigma_K\cr 0 &\Sigma_A\cr\end{pmatrix},
\cr\nonumber\\&&
{ \tilde S}=\begin{pmatrix}{\tilde S}_R &{\tilde S}_K\cr 0 &{\tilde S}_A\cr\end{pmatrix}        
\end{eqnarray}
or in components,
\begin{eqnarray}\label{ds1}
&&
{\tilde S}_R=S_R+iS_R*\Sigma_R*{\tilde S}_R
\cr\nonumber\\&&
{\tilde S}_A=S_A+iS_A*\Sigma_A*{\tilde S}_A
\cr
\nonumber\\&&
{\tilde S}_K=S_K+i[S_R*\Sigma_K*{\tilde S}_A
\cr\nonumber\\&&
+S_K*\Sigma_A*{\tilde S}_A+S_R*\Sigma_R*{\tilde S}_K].
\end{eqnarray}
{The formal} solution is
\begin{eqnarray}\label{ds2}
&&
{\tilde S}_R=[1-iS_R*\Sigma_R]^{-1}* S_R=S_R*[1-i\Sigma_R*S_R]^{-1},
\cr\nonumber\\&&
{\tilde S}_A=[1-iS_A*\Sigma_A]^{-1}* S_A=S_A*[1-i\Sigma_A*S_A]^{-1},
\cr\nonumber\\&&
\tilde S_K=+i\tilde S_R\Sigma_K\tilde S_A
+\tilde S_R S_R^{-1}S_K S_A^{-1}\tilde S_A
\end{eqnarray}
\begin{eqnarray}\label{ds3}
&&{\tilde S}_K=-S_{K,A}*(1-i\Sigma_A*S_A)^{-1}
+(1-iS_R*\Sigma_R)^{-1}*S_{K,R}
\cr\nonumber\\&&
+i(1-iS_R*\Sigma_R)^{-1}*[S_R*\Sigma_K*S_A
-S_R*\Sigma_R*S_{K,A}
\cr\nonumber\\&&
+S_{K,R}*\Sigma_A*S_A]*(1-i\Sigma_A*S_A)^{-1},
\end{eqnarray}
where we have~used 
\begin{eqnarray}\label{ds4}
&&
S_K=S_{K,R}-S_{K,A},~~
 \Sigma_K=\Sigma_{K,R}-\Sigma_{K,A}.
\end{eqnarray}

\subsection{Fermion Particle~Number}
\label{FermionParticleNumber}

The number of fermions of momentum $\vec p$ at the time $t$ is
obtained from the time evolution of the of number operator, which is 
the Average of the Equal Time Limit (AETL) of the propagator $S_{K,s_{01},s_{02}}$.
The average is necessary because we must distinguish the limit from above 
$s_{01}-s_{02}=\Delta>0$ 
and the limit from below $\Delta<0$. If~the propagator $\tilde S_{K,s_{01},s_{02}}(p)$ 
has a pole which is apart from the real axis for the finite imaginary part, these two limits 
will be different, in which case we take the average. Using~AETL, we are able to 
understand the limit:
\begin{eqnarray}\label{AETL}
&&
AETL={1\over 2}[\lim_{0<\Delta \rightarrow 0}+\lim_{0>\Delta \rightarrow 0}].
\end{eqnarray}

The particle number is defined~as

\begin{eqnarray}\label{nude}
&&  \langle \, N_{\vec p}(t)\, \rangle  \, = \, (2\pi)^3 \, d{\cal N}/(d^3xd^3p)
\end{eqnarray}
and 
\begin{eqnarray}\label{nupaf}
&&
 1-\langle \,N_{f, \vec p}(t)  \, \rangle  \,
= \, {1\over 2\pi}[\lim_{0<\Delta \rightarrow 0}
+\lim_{0>\Delta \rightarrow 0}]\int dp_0 
	e^{-ip_0\Delta} \, Tr [\, {\gamma_0\over 4}\tilde S_{K,t}(p) \, ],
\end{eqnarray}
where $\Delta=s_{01}-s_{02}$ and $X_0=(s_{01}+s_{02})/2=t$.

The contribution (\ref{nupaf}) corresponds to a single polarization.
To sum over the polarizations, an additional factor of $2$ is~necessary.

When acting on the momentum and spin eigenstates of fermions $|+,\vec p,s \, \rangle$
and antifermions $|-,-\vec p,s \, \rangle $ (both normalized to
$\langle \, \pm, \pm \vec p,s|\pm, \pm \vec p,s \, \rangle \, = \, 1/m $),
the projectors $\Lambda_{\pm}$ satisfy

$\Lambda_{+}(\omega_p,\vec p)|+,\vec p,s  \, \rangle \, = \, |+,\vec p,s  \, \rangle \quad $
 and
$\quad  \Lambda_{-}(\omega_p,\vec p)|+,\vec p,s \, \rangle \,  = \, 0~, \quad\quad $ while

$\Lambda_{+}(\omega_p,\vec p)|-,-\vec p,s \, \rangle \, = \, 0 \quad $ and
$\quad  \Lambda_{-}(\omega_p,\vec p)|-,-\vec p,s \, \rangle \, = \, |-,-\vec p,s\, \rangle$.

In the rest of this paper, the~distribution functions are
assumed to depend on $|\vec p|$ and~not on the direction of $\vec p$.
The lowest order contribution ($N^0_{f+\bar f,\vec p}(t)$)
 is identical to the initial distribution (for $n_{\bar f}(\omega_p)=0$):
\begin{eqnarray}\label{nupa0n}
1 - \langle \, N^0_{f,\vec p}(t) \, \rangle \, = \, 1 - n_{+f}(\omega_p)~.
\end{eqnarray}

\subsection{Massless Chiral Fermion Particle~Number}
\label{A8}

For a massless {\it chiral} particle there is one more helicity projector 
$(1-\gamma_5)/2$ with respect to the trace in Equation~(\ref{nupaf}).
The particle number becomes (\ref{nupaff}):
\begin{eqnarray}\label{Anupaff}
	1 &-& \langle \,N_{\beta, \vec p}(t)  \, \rangle  \,
\cr\nonumber\\
&=&{1\over 2\pi} \, [\, \lim_{0<\Delta \rightarrow 0}
+\lim_{0>\Delta \rightarrow 0}\, ]\int dp_0  \, e^{-ip_0\Delta} \, Tr [\,
{\gamma_0\over 4} \, {\tilde S}_{\beta,\beta,K,t}(p) \, {1-\gamma_5\over 2} \, ] \, .
\end{eqnarray}

Here, $\langle \,N_{\beta, \vec p}(t)  \, \rangle$ is the number
of neutrinos of flavor $ f$ and momentum $\vec p$
detected at time $t$. 

Note that one more helicity projector ${1-\gamma_5\over 2}$ is
necessary to ensure the proper helicity (chirality) of the detected neutrinos.
The lowest-order contribution for a massless chiral particle 
 is identical to the initial distribution: 
\begin{eqnarray}\label{nupa0f}
1- \langle N^0_{f\vec p}(t) \rangle \, = \, 1-n_{f}(|\vec p|) \, .
\end{eqnarray}

\begin{adjustwidth}{-\extralength}{0cm}

\reftitle{References}

%=====================================

\PublishersNote{}
\end{adjustwidth}
\end{document}